# Monitoring cis-to-trans isomerization of azobenzene using Brillouin microscopy

Zhe Wang[1], Qiyang Jiang[1], Chantal Barwig[1], Ankit Mishra[1], Krishna Ramesh[1,2] and Christine Selhuber-Unkel[1,2]

[1] Institute for Molecular Systems Engineering and Advanced Materials (IMSEAM), INF 225, Heidelberg University, 69120 Heidelberg, Germany.
[2] Max Planck School Matter to Life, Jahnstraße 29, 69120 Heidelberg, Germany.

E-mail: selhuber@uni-heidelberg.de

**Abstract**

Brillouin spectroscopy is commonly used to study the acoustic properties of materials. Here we explored its feasibility in studying the photoinduced isomerization of azobenzene. The isomerization of azobenzene changes the solution elastic modulus, and Brillouin scattering is sensitive to these changes. In this study, we experimentally demonstrated the photoswitching of azobenzene in DMSO using our home-made VIPA-based high-resolution optical Brillouin spectrometer, and confirmed the results by ultraviolet–visible spectrophotometry. Remarkable Brillouin frequency shift variations were quantitatively recorded upon irradiation, and it was found that this method can indeed be used to monitor the isomerization process *in situ*. Importantly, our strategy also allows us to provide the relationship between the fraction of trans- and cis- azobenzene and the Brillouin frequency shift. This shows that Brillouin spectroscopy has broad prospects for the characterization of azobenzene isomerization and other photo responsive materials.

## 1. Introduction

Azobenzene is one of the most promising photo-switchable molecules that undergoes conformational and polarity changes between trans- and cis-isomers upon isomerization of the central diazene double bond [1-5]. The thermodynamically stable trans-isomer can be switched from trans to cis upon illumination with ultraviolet (UV) light [6]. The reverse process from the metastable cis-isomer to the trans-isomer can be triggered by either light illumination (visible light) or thermal relaxation [7]. This photoswitchable property endows azobenzene-based nanoscale molecular machines with the capability to function as unidirectional motors with exclusive forward rotary motion or as a back-and-forth molecular switch [8, 9]. In addition, this extreme tunability demonstrates unique superiority for the production of photoswitchable microrobots [10] and biomaterials, especially for the study of cellular mechanotransduction in the *in vitro* and *in vivo* microenvironment [11, 12]. To date, azobenzene-based photo stimulation of cells has mostly been investigated by atomic force microscopy (AFM) and micropipette aspiration to reveal mechanistic details of cell system machinery [13-15]. For example, a push-pull azobenzene modified integrin binding RGD peptide has been demonstrated to achieve rapid and reversible cell adhesion switching using single-cell force spectroscopy(SCFS) [15].

However, the conversion of the trans- and cis-isomers by light is not 100%. Azobenzenes reach a photostationary state (PSS), which means that extending the exposure to light does not change the conversion ratio. The fraction of trans- and cis-isomers depends on many factors, such as molecular structure and temperature [16]. It is typically calibrated using UV-Vis spectrophotometry and ¹H-NMR spectroscopy. More recently, terahertz time-domain spectroscopy has also been employed for this purpose [7]. Pump-degenerate four wave mixing has also shown the potential by mapping the evolution of fingerprint vibrational modes during isomerization [17, 18]. For the above approaches, a special setup is required, and it would be extremely difficult to use for monitoring azobenzene isomerization *in situ* during biological experiments. In addition, the absorption peak of azobenzene is in the UV range for UV-Vis spectrophotometry, potentially inducing a change in the sample state during probing. Thus, Brillouin microscopy has emerged as a unique technique especially as it can be installed as an add-on Brillouin module to most commercial microscopes, capable of being pumped with any wavelength [19-22].

Brillouin microscopy is a valuable method for quantitatively assessing the mechanical characteristics of materials. It is a non-invasive, label-free, and high-resolution mechanical imaging technique with minimal interference and can be used to measure mechanical anisotropy [23-26]. Nevertheless, its application in microscope has been limited until recently owing to its weak signals [27, 28]. The enhancement of Brillouin scattering signal represents a key area of interest in this field. For example, stimulated Brillouin scattering (SBS) process could significantly enhance the signal-to-noise ratio through substantial amplification [29, 30].

In this work, the measurement is based on spontaneous Brillouin scattering, which is inelastic light scattering with phonons or thermally populated acoustic waves. The excitation lights are generally shifted in direction and frequency after coupling with phonons or sound waves. The Brillouin frequency shift $v_b$ is directly related to the speed of sound in the material and the angle of the scattered light, which is given by:

$$v_b = \frac{2n}{\lambda} u \sin\left(\frac{\theta}{2}\right), \tag{1}$$

where $n$ is the refractive index, $\lambda$ is the wavelength, $u$ is the acoustic velocity and $\theta$ is the scattering angle [31]. The medium elastic modulus $M$ can also be related through the relation $M(v_b) = \rho u^2$, where $\rho$ is the density. Thus, this technique can be applied to characterize the acoustic viscosity and apparent elastic modulus [32]. In the case of the push-pull azobenzene, alterations in the molecular geometry during isomerization reactions can impact the interactions between azobenzene molecules and the surrounding solvent molecules [33-35]. Therefore, Brillouin spectroscopy offers a promising method to track these isomerization reactions in real time.

In this work, we synthesized a new ortho-fluorinated azobenzene which offers favorable photochemical characteristics such as long thermal stability and efficient photoisomerization. Its long thermal stability allows orthogonal and reversible control of the isomerization process by using two different wavelengths. This is particularly advantageous for the precise quantification of the cis-to-trans conversion that is initiated upon exposure to light. We employed the Brillouin microscopy technique to observe the reversible cis-to-trans isomerization of Azo, marking the first known application of the technology in this particular field.

## 2. Experimental

*Light trigger of Azobenzene*

The azobenzene shown in Fig. 1, substituted by a weak electron donating group (EDG) and electron-withdrawing group (EWG) was selected and synthesized in three steps by modifying a previously described procedure and is abbreviated as Azo in this work, as detailed in the SM. A solution of Azo (0.05 mM) was freshly dissolved in DMSO and introduced into a 400-μL quartz cuvette with a light path of 10 mm (Hellma 104-002-10-40), which was illuminated under 365 nm (LED: 8 x Nichia NVSU233A-U365) and 565 nm (LED:3 x Luxeon-LXML-PX02) with increased exposure time. At selected exposure times, the UV-Vis spectrum of the solution was recorded. The power of the LEDs was measured using a power meter (PM100, Thorlabs).

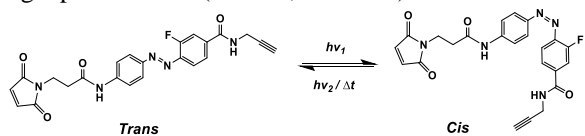

Figure 1. Light-triggered transition of trans- and cis-Azo structures.

*UV-Vis spectrophotometer*

UV-Vis absorbance spectra were measured with a spectrophotometer (JASKO V-770) equipped with a single monochromator using 120 lines/mm for the UV/Vis region and a PMT detector. In this instrument, a monochromator was employed to produce unpolarized light at a specified wavelength. The generated light was subsequently divided into two pathways: one passing through the sample, and the other serving as a reference. The transmission and absorbance of the sample can be calibrated by comparing the light detected in the sample branch with that detected in the reference branch in real-time.

*Brillouin spectroscopy*

Brillouin scattering measurements were performed with a virtually imaged phased array (VIPA) -based high-resolution optical Brillouin spectrometer as shown in Fig. 2 [22, 36]. The setup consists of a commercially available inverted confocal microscope (IX81, Olympus) coupled with a home-made Brillouin module. This module integrates a continuous-wave laser (532 nm, Cobolt Samba 100), an external optical setup for directing light into and out of the microscope, and a Brillouin spectrometer.

To briefly describe the optical setup, after passing through an optical isolator, the input laser is reflected by a polarized beam splitter (PBS) and guided to the microscope. It is then focused by a 20x objective on the sample with a beam expander to overfill its back focal aperture. After the sample, the backward-scattered Brillouin light is collected by the same objective and its polarization is rotated 90 degree by a quarter-wave plate (QWP) such that it can be separated by the PBS. Finally, the signal is collected by a single-mode optical fiber to the Brillouin microscope, which also makes it confocal.

The detail of our Brillouin microscope can be found in Jitao Zhang et al. and Kim V. Berghaus et al. [22, 36]. It is composed of two orthogonally oriented VIPA stages, two spatial filters (masks) and a detection system with a 4-f unit and sCMOS camera (PCO, Edge 4.2). The advantage of cross-axis VIPA stages is that we can spatially filter out Rayleigh scattering and its crosstalk substantially. This enabled us to complete the experiments in a short time, which is crucial for optical switching experiments.

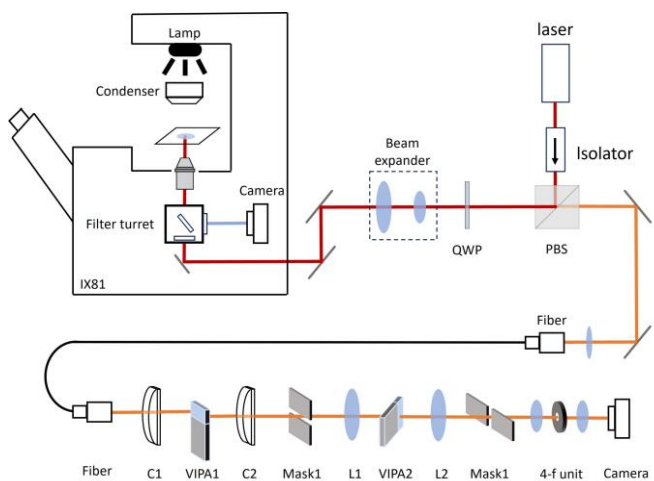

Figure 2. Schematic of the Brillouin microscope setup. PBS, polarized beam splitter; QWP, quarter-wave plate; C1, C2, cylindrical lenses; VIPA1, VIPA2, virtually imaged phased array; L1, L2, lens.

## 3. Results and disscussion

### UV–Vis absorbance spectroscopy

To quantify the concentration of trans- and cis- Azo in the DMSO solution, its absorbance under different conditions was first studied by UV-Vis spectroscopy. The UV-Vis-spectrum of Azo in DMSO ($c = 0.05$ mM) showed an absorption maximum at $\lambda_{max} = 377$ nm with a broad shoulder at $\lambda \sim$ 440-500 nm (Fig. 3a). As shown in Fig. 3a, upon irradiation at 365 nm, the intensity of the maximum absorption continuously decreased to reach a PSS, after 90 s of irradiation. No further changes were observed in the spectra after 90 s irradiation. The photoisomerized cis-isomer fraction of Azo molecules (referred to as cis fraction in this paper) with UV was determined to be $73.1 \pm 0.2\%$. Therefore, we refer to this state as $PSS_{cis}$. A similar PSS state with $66.0 \pm 0.3\%$ trans-Azo could also be achieved with 90 s 565 nm LED irradiation, which is called $PSS_{trans}$ here and shown in Fig. S2. The fraction was determined by measureing the absorbances at 377 nm using the following equation:

$$R_{cis}[\%] = \frac{(A_o - A_{PSS})}{A_o} \times 100\% \quad (2)$$

where $A_o$ is the absorbance at $\lambda_{max}$ before irradiation (100% trans-Azo) and $A_{PSS}$ is the absorbance of PSS after irradiation, as shown in Fig. 3b. Here, we assume that the Beer-Lambert law holds,

$$A(\lambda) = [c_{cis} \times \varepsilon_{cis}(\lambda) + c_{trans} \times \varepsilon_{trans}(\lambda)] \times L \quad (3)$$

where $A$ is the measured wavelength-dependent absorbance, $\varepsilon$ is the molar extinction coefficient of the trans or cis state Azo, c is the concentration and L is the optical path length [37]. The cis fraction was further verified using $^1$H-NMR-spectroscopy. The corresponding integrals revealed a percentage of cis-isomer in PSS of $70 \pm 3\%$), which matches the results from UV-Vis-investigation ($R_{cis} = 73.1 \pm 0.2\%$); details are summarized in the Supporting Information (Fig. S1). According to the above results, the molar absorption coefficients ($\varepsilon$) of trans-Azo and cis-Azo were also calibrated as shown in the inset of Fig 3b, which is consistent with previously reported data [37]. Furthermore, the absorption peaks corresponding to the photoswitching process were approximately 377 nm (for trans to cis) and 440 nm (for cis to trans), as illustrated in the inset of Fig. 3b. We selected 565 nm LEDs because of their broad spectrum. It is noteworthy that employing a 440 nm LED would encompass a wavelength of 377 nm within its spectral range.

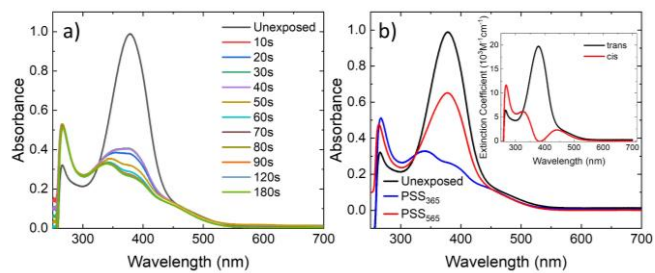

Figure 3. (a) UV-Vis absorbance spectra of Azo in solutions (c = 0.05 mM in DMSO) after UV light (centered 365 nm, $5.0 \pm 0.2$ mW cm$^{-2}$) over different time spans; (b) UV-Vis absorbance spectra after UV light (centered 365 nm, $5.0 \pm 0.2$ mW cm$^{-2}$) and green light (centered 565 nm, $5.0 \pm 0.2$ mW cm$^{-2}$) irradiation with 180s exposure time; the inset is the molar absorption coefficients ($\varepsilon$) of trans-Azo and cis-Azo at room temperature.

*The Brillouin shift*

To calibrate our Brillouin microscopy, two reference materials (water and DMSO) were measured at the sample position using a $22.0 \pm 0.7$ mW laser power, as illustrated in Fig. 4a and 4b, respectively. In all experiments, we maintained a constant probe laser power to ensure consistent conditions, achieving a signal-to-noise ratio (SNR) of 26.8 dB (in the case of DMSO) [29, 38]. Without spatial filtering, there are multiple dispersion orders of the spectral patterns (Rayleigh scattering, Stokes, and anti-Stokes Brillouin scattering). Because the Rayleigh scattering intensity is much stronger than the Brillouin scattering intensity, the camera can be easily saturated. Thus, Rayleigh scattering was blocked by mask1 and mask2. The bright dots shown in the center of Fig. 4a and 4b are anti-Stokes Brillouin peak from one dispersion order and the Stokes Brillouin peak from the subsequent order. The spectra extracted from them (light intensity along the gray line in a and same location in b ) and the corresponding Lorentzian fittings are shown in Fig. 4c. The distinct peaks observed were quantified in pixels, and the pixel-to-frequency conversion ratio (PR= $0.02418 \pm 0.00007$ Ghz/pixel) and frequency spectral range (FSP= $22.06 \pm 0.02$ GHz) were calibrated according to them [22, 36].

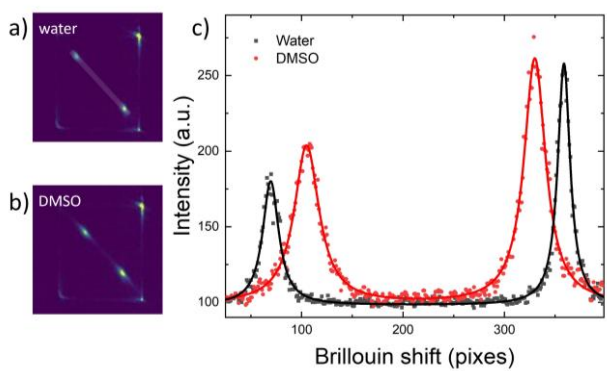

Figure 4. (a) and (b) Acquired Brillouin spectra of water and DMSO by sCMOS camera, which are used for calibration. (c) Lorentzian curve fitted to the measured data according to the gray line in (a) and the same position in (b).

*Photoinduced isomerization of Azo*

The photoinduced trans-cis isomerization of Azo irradiated with LED light (centered at 365 nm and 565 nm) in DMSO at room temperature was studied. We exposed the sample for 3 min with LED ($5.0 \pm 0.2$ mWcm$^{-2}$) to ensure the Azo was at photostationary state. Again, the "trans" configurations (PSS$_{trans}$) are illuminated using LED centered at a wavelength of 565 nm while the "cis" configurations (PSS$_{cis}$) were illuminated using LED centered at a wavelength of 365 nm. After light exposure, the camera captured a series of 10 frames, each with an exposure time of 500 ms to reduce the shot noise. It is noticed that laser wavelength is 532 nm, which is close to the critical switching wavelength. To mitigate the influence of the laser on the experiment, the LED stayed on for the duration of the measurements. The inset of Fig. 5 presents the Brillouin frequency shifts associated with the last data point, which verifies the stability of the Azo compound and demonstrates that the presence of laser did not significantly affect our observations.

To demonstrate the reversibility of Azo photoswitch, we conducted five rounds of photoswitching between PSS$_{trans}$ and PSS$_{cis}$, with the transitions monitored using our Brillouin microscope. The results in Fig. 5 enable both qualitative and quantitative validation of the true reversibility of Azo photoswitching and show the prospect of using Brillouin microscopy to observe the Azo PSS. Brillouin frequency fluctuations were observed, with a decrease corresponding to UV radiation exposure and an increase during visible light exposure. These results indicate that the isomerization-induced changes in molecular geometry can affect the interactions between Azo molecules and the surrounding solvent molecules. In addition, the altered shape and size of the molecule may lead to changes in the surrounding solvation shell, which manifest as altered mechanical changes in the elastic modulus [33-35, 39].

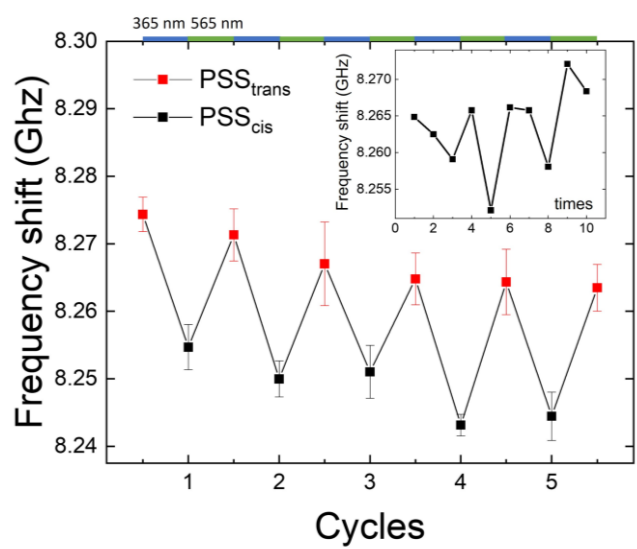

Figure 5. Average Brillouin frequency shift of Azo over five cycles of photoswitching with LED centered at a wavelength of 365 nm and 565 nm due to the trans-cis isomerization. Error bars: Standard error.

This method can be utilized to quantify the fraction of trans- and cis- Azobenzene in solution after calibration. Here we used two PSS as a reference and showed the cis-Azo ratio change during the exposure. A formal biphasic model was applied with trans-Azo in DMSO as the first phase and cis-Azo in DMSO as the second phase. The consistent correlation between the Brillouin line frequency and mass fraction across most of the studied samples supports an explanation free from chemical substance dependence[40]. This implies that a portion of the elastic modulus, denoted as $M$, can be attributed to the solutions. Under the same stress conditions, the effective elastic modulus of the solution, denoted as $M_{\text{eff}}$, is expressed as:

$$\frac{1}{M_{\text{eff}}} = \frac{x_{\text{trans}}}{M_{\text{trans}}} + \frac{x_{\text{cis}}}{M_{\text{cis}}} \quad (4)$$

where $M_{\text{trans}}$ and $M_{\text{cis}}$ are the elastic modulus of trans Azo and cis Azo solutions, respectively, while $x_{\text{trans}}$ and $x_{\text{cis}}$ are their fraction of Azo molecules, with $x_{\text{trans}} + x_{\text{cis}} = 1$[41]. Similar to the reference, we also neglect variations in the refractive index with changes in the concentrations of trans and cis Azo, as well as any alterations in the solution density. Together with Equation 1, the effective Brillouin frequency $v_{\text{b}}^{\text{eff}}$ of the solution can be expressed as:

$$v_{\text{b}}^{\text{eff}} = \frac{v_{\text{b}}^{\text{trans}}}{\sqrt{1 - x_{\text{cis}} + \frac{x_{\text{cis}} u_{\text{trans}}^2}{u_{\text{cis}}^2}}} \quad (5)$$

where $v_{\text{b}}^{\text{trans}}$ is the Brillouin frequency of the trans-Azo solutions, $u_{\text{trans}}$ and $u_{\text{cis}}$ are the sound velocities of the trans-Azo and cis-Azo solution, respectively. According to the Azo fraction of PSS$_{\text{trans}}$ (66.0±0.3% trans-Azo) and PSS$_{\text{cis}}$ (73.1±0.2% cis-Azo) by UV-Vis absorbance spectra, we can estimate the Brillouin frequency of trans and cis Azo with $v_{\text{PSS\_trans}}^{\text{eff}} = 8.264 \pm 0.004$ GHz, $v_{\text{PSS\_cis}}^{\text{eff}} = 8.244 \pm 0.003$ GHz. It is worth mentioning that since the measurements from the initial cycles showed variability, our analysis was conducted using the average from the last three cycles for consistency. A similar behaviour was oberved in the UV-Vis absorbance spectra, (Fig. S3). In conclusion, the Brillouin frequency $v_{\text{b}}^{\text{trans}}$ is $8.277 \pm 0.008$ GHz and $v_{\text{b}}^{\text{cis}}$ is $8.226 \pm 0.006$ GHz.

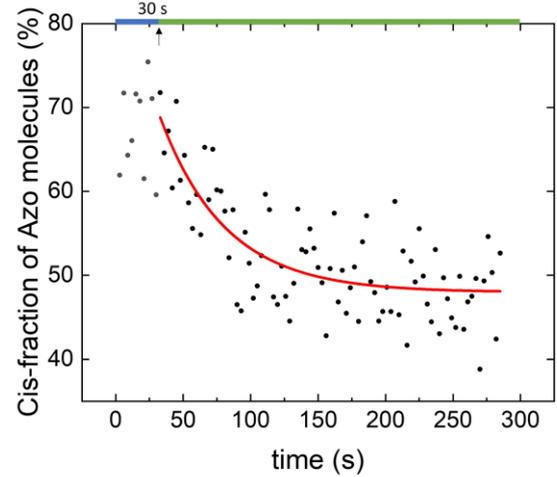

Fig.6. Estimated cis-fraction of Azo molecules from PSS$_{\text{cis}}$ to PSS$_{\text{trans}}$. The Azo sample was at PSS$_{\text{cis}}$ while the 365 nm LED was switched to 565 nm at 30 s. The red line is fitted by exponential decaying function.

Based on the obtained results, we were able to determine the changes in the fraction that occurred during the photoinduced isomerization reactions, as shown in Fig. 6. The Azo solution was exposed for 3.5 min under a 365 nm LED to reach PSS$_{\text{trans}}$, similar to previous experiments. Brillouin shifts were recorded 30 s before the 565 nm LED, and the cis-fraction of Azo molecules was estimated. The observed trend showed an initial decrease in the cis-fraction, which then asymptotically settled towards a different photo-stationary state. In addition, an exponential decay function was applied for fitting and a time constant of $50 \pm 10$ s was obtained.

## 4.Conclusions

In this work, we have first demonstrated the quantitative monitoring of the photoinduced isomerization processes of Azo using Brillouin microscopy. The relationship between the fraction of trans- and cis- Azo and the Brillouin frequency shift is also provided. Brillouin microscopy facilitates the precise distinction and quantification of the conversion state between isomers in response to light, providing novel insights into the isomerization processes. Leveraging the exceptional sensitivity of this technique, we anticipate that decisive and nondestructive Brillouin microscopy holds promising avenues for investigation, leading to a deeper understanding of the conformational dynamics of molecules.

## Acknowledgements

Funding from the European Research Council (ERC CoG no. 101001797 PHOTOMECH) is acknowledged. C.S. and K. R. are supported through the MPS Matter to Life supported by the BMBF. We also thank the Flagship Initiative "Engineering

Molecular Systems" funded by BMBF and the Ministry of Science Baden- Württemberg within the framework of the Excellence Strategy of Federal and State Governments of Germany. The authors also acknowledge funding by the DFG under Germany's Excellence Strategy 2082/1-390761711 (3D Matter Made to Order), as well as through the RTG 2154 and the project SE1801/4-2 within the Priority Program SPP 2206 "KOMMMA". The Brillouin microscope was a kind gift from Thomas Greb's group. The authors thank Robert Prevedel, Carlo Bevilacqua and Theresa Schlamp for their support with alignment and discussion. The support by the NMR laboratory of the Anorganisch-Chemisches Institut, Heidelberg University, is gratefully acknowledged. The authors acknowledge the support of Dr. Maria Regato Herbella.

**Conflict of interest**

The authors have declared that no competing interests exist.